\documentclass{XrU2005}
\usepackage{epsfig}

\title{X-ray Reflections on AGN}
\author{A.C. Fabian}
\affil{Institute for Astronomy, Madingley Road, Cambridge CB3 0HA, UK}

\begin{document}

\keywords{}

\maketitle

\begin{abstract} X-ray reflection generates much of the spectral
complexity in the X-ray spectra of AGN. It is argued that strong
relativistic blurring of the reflection spectrum should commonly be
expected from objects accreting at a high Eddington rate. The good
agreement found between the local density in massive black holes and
the energy density in quasar and AGN light requires that the accretion
which built massive black holes was radiatively efficient, involving
thin discs extending within 6 gravitational radii. The soft excess
found in the spectra of many AGN can be explained by X-ray reflection
when such blurring is included in the spectral analysis. Some of the
continuum variability and in particular the puzzling variability of
the broad iron line can be explained by the strong light bending
expected in the region immediately around a black hole. Progress in
understanding this behaviour in the brightest sources can be made now
with long observations using instruments on XMM-Newton and Suzaku.
Future missions like Xeus and Con-X, with large collecting areas, are
required to expand the range of accessible objects and to make
reverberation studies possible.

\end{abstract}

\section{Introduction} 

In this brief review, I consider the spectra and spectral variability
of unobscured Active Galactic Nuclei such as Seyferts and quasars.
They typically have the spectral components identified in Fig.~1,
namely a) an underlying power-law, b) a soft excess above the
power-law at low energies below 1~keV, c) an iron line (which may have
a broad component), and d) a Compton hump. Traditionally these
components have been considered as a) thermally Comptonized soft
photons originating from b) thermal (blackbody) emission from an
optically-thick accretion disc about the central black hole, together
with the line c) and Compton-scattered d) parts of X-ray reflection
from that disc or more distant matter. An important parameter when
model-fitting such sources is the inner radius of the accretion
disc, which determines how much relativistic blurring is applied to
the reflection components.  It is often assumed to be greater than 6
gravitational radii ($6 r_{\rm g}=6GM/c^2$) around the black hole,
which is the innermost stable circular orbit around a non-spinning
Schwarzschild black hole. Spectral deviations from this picture are
often taken into account by adding additional emission and/or absorption
components, some of which cover only part of the source.

\begin{figure}[h]
\centering
\includegraphics[width=0.65\linewidth,angle=270]{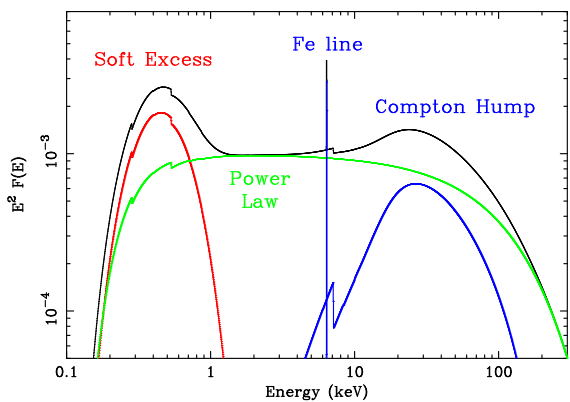}
\caption{Model X-ray spectrum of an AGN. Galactic
absorption causes the flux to decrease steeply below 0.3~keV.}
\end{figure}

There are problems with this traditional picture which suggest
that it is at least incomplete. For several well-studied sources there
are modifications to the above model which seem to fit the data, in
particular the spectral variability, better. The main modification is
to allow the inner radius of the disc to go to $2r_{\rm g},$ meaning
the black hole is spinning. This introduces the possibility of very strong
gravitational effects on the spectrum. The second consideration is to
allow the atomic abundances to be different from the solar value.
A more detailed review is given in Fabian \& Miniutti (2005).

Note that most bright AGN {\em must} have a radiatively efficient
accretion flow or the Soltan (1982) argument relating the energy
density of accretion radiation and the local mean density in black
holes would yield a low efficiency. The good agreement between the
observations of quasar/Seyfert light and local black holes with an
accretion efficiency of at least 10 per cent (Yu \& Tremaine 2002;
Fabian 2003; Marconi et al 2004) strongly argues for radiatively
efficient flows with an inner disc radius within $6r_{\rm g}$. The
agreement, at all mass ranges, would not happen if the discs in
quasars and luminous Seyfert galaxies stopped at several tens of
$r_{\rm g}$ or indeed larger than $6r_{\rm g}$. Any power lost in
winds and jets only strengthens these arguments. Massive black holes
in galactic nuclei are likely to be rapidly spinning (Volonteri et al
2004) so small disc inner radii should be the norm and we should
seriously consider that much of the X-ray emission from objects
accreting at a high ($\gg0.01$) Eddington fraction emerges from within
a few $r_{\rm g}$.

\section{The problems}

\subsection{The soft excess}

Several studies culminating in the work of Gierlinski \& Done (2004)
show that the temperature of the excess emission, if characterized as
blackbody, seems to be the same in systems where the accretion rates
and/or masses differ by several orders of magnitude.  This is not
expected from an accretion disc.

\subsection{The iron line}		

Many sources show a narrow iron line component which is undoubtedly
due in many cases to reflection on distant gas. Broad components, as
expected from reflection by the inner accretion disc, are
seen, but are not always present or at least not
evident. Such components can sometimes be fitted away with
partial-covering models.  

\subsection{Variability}

Where sources are highly variable so that the emission region must be
very small, partial covering models present physical problems for
understanding the geometry of the situation. Only very
occasionally can we be in a preferred line of sight;  the covering material
has to be randomly placed. What this matter is, where it lies and why
it only partly covers the source are  unknown.

\begin{figure}[h]
\centering
\includegraphics[width=0.65\linewidth,angle=270]{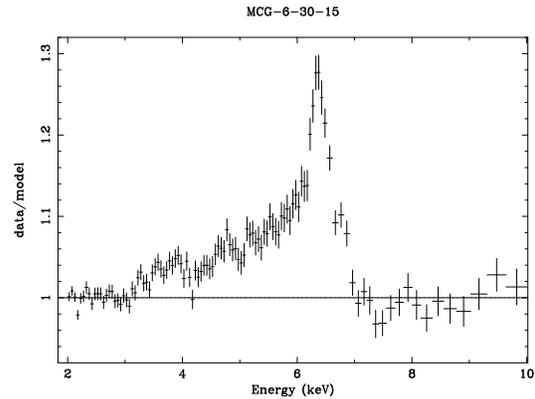}
\caption{The broad iron line seen in the XMM-Newton spectrum of  
MCG--6-30-15 (see Fabian et al 2002; this spectrum was produced from
reprocessed data by S. Vaughan).}
\end{figure}
\begin{figure}[h]
\centering
\includegraphics[width=0.65\linewidth,angle=270]{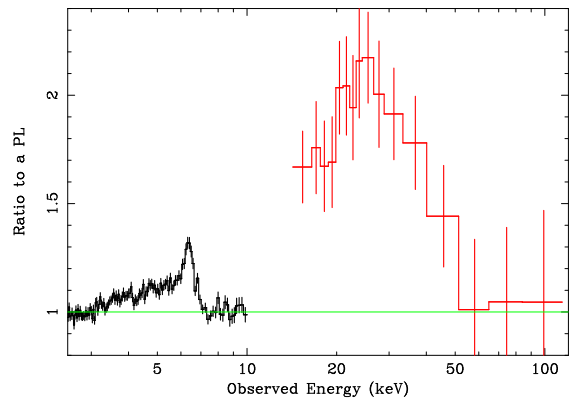}
\caption{The iron line and Compton hump in MCG--6-30-15 shown as
deviations from a simple power-law spectrum. The data below 10~keV are
from XMM-Newton and the data above from BeppoSAX (kindly prepared by
G. Miniutti).}
\end{figure}
\begin{figure}[h]
\centering
\includegraphics[width=0.65\linewidth,angle=270]{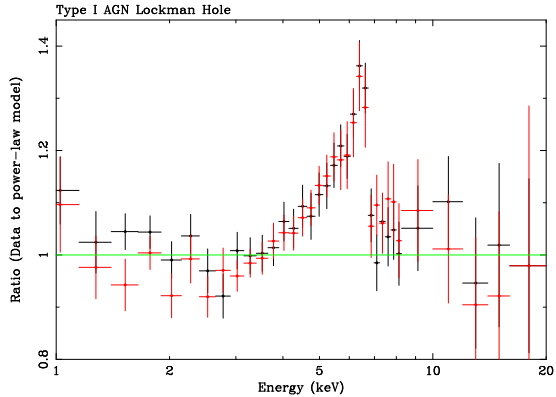}
\caption{Broad iron line seen in the rest-frame shifted, summed 
spectra of 51 Seyfert 1 galaxies in the XMM-Newton observations 
of the Lockman Hole (Streblyanska et al 2005). }
\end{figure}

\subsection{Iron line variability}

MCG--6-30-15 has a robust broad iron line (Tanaka et al 1995; Wilms et
al 2001; Fabian et al 2002). Chandra grating
observations and RXTE data have sufficient resolution and coverage to
rule out partial covering solutions (Young et al 2005). A problem
emerges with the lack of variability seen in the line, if the effects
of strong gravity are ignored. The strength
of the iron line should follow the brightness of the power-law
component, but it does not. The iron line does vary on short
timescales but not in any simple manner (Iwasawa et al 1996, 1999;
Nandra \& Edelson 2000; Matsumoto et al 2002; Fabian et al 2002).

\section{The two-component model of spectral variability}

A simple phenomenological model which fits the spectral variability of
several bright Seyferts well has two main components (McHardy et al
1998; Shih et al 2001; Fabian \& Vaughan 2003). A simple power-law
decription of the spectrum often shows the source to be harder when
faint and softer when bright. The photon index of the source may limit
to some fixed value at the highest fluxes. This behaviour can be
modelled well in terms of two components; a soft power-law of fixed
spectral index and variable intensity plus a hard component which
varies little. The model accounts for  the rms variability spectrum
and spectral behaviour of many sources.

\begin{figure}[h]
\centering
\includegraphics[width=0.65\linewidth,angle=270]{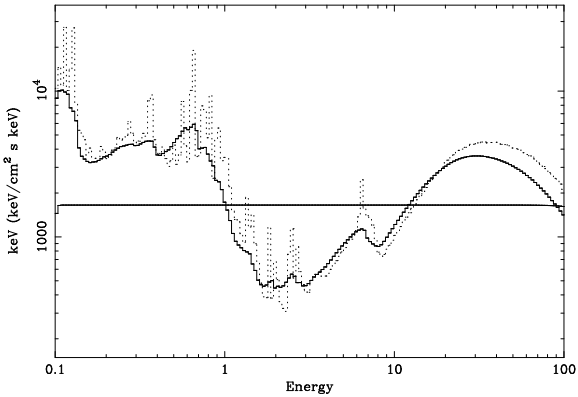}
\caption{Two-component model. The horizontal line is the power-law
component which varies significantly in amplitude. The solid curved
line is the relativistically-blurred reflection spectrum (shown dotted).}
\end{figure}

The shape of the quasi-constant hard component can be extracted in
several ways, using a) flux-flux plots in which the flux in various
energy bands is correlated with the flux in another band (say
1--2~keV) with the constant component appearing as the intercept (e.g.
Taylor et al 2004); b) difference plots where the spectrum of the
faintest flux state is subtracted from that of the brightest (Fabian
et al 2002); or c) straight model fitting (Vaughan et al 2004). When
applied to MCG--6-30-15 all these methods indicate that the hard
component has the shape of blurred reflection and the variable soft
component is a power-law.  Similar results are found for NGC\,4051
(Ponti et al 2005).

\begin{figure}[h]
\centering
\includegraphics[width=0.65\linewidth,angle=270]{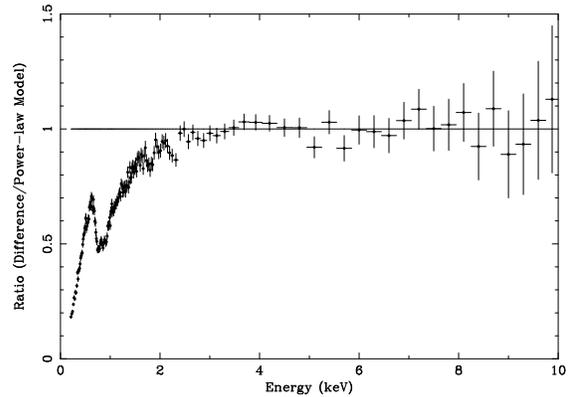}
\caption{Difference spectrum from MCG--6-30-15 (Turner et al
2004). It shows the ratio of the spectrum, made by subtracting low from
high flux data, to a simple power-law. Note that it has no broad
iron line and shows that the component which varies is a
simple power-law in the 3--10~keV band). Assuming that it remains a
power-law to low energies, the deviations there show the absorption
components (mostly due to a warm absorber).} 
\end{figure}

\section{Solutions invoving strong gravity}

An alternative interpretation of the soft excess, hinted at in earlier
papers (Czerny et al 2003; Ross \& Fabian 2005) is for it to be the
blurred soft part of the ionized reflection spectrum. This has been
tested by Crummy et al (2005) and generally found to give better fits
to spectra of PG quasars and various Seyferts than a simple blackbody
disc does. The reflection spectrum needs to be significantly blurred,
requiring that much of the emission arises from near the centre of an
accretion disc about a spinning black hole. The blurred reflection
spectrum has a 'boxy' shape better suited to the soft excess than a
blackbody which then requires fewer, if any, additional absorption components
for a good fit.

The iron abundance needs also to be a free parameter. Extreme blurring
together with low iron abundance can make most broad iron lines
undetectable with current instruments. 

If much of the X-ray emission emerges from the innermost parts of the
disc around a spinning black hole then light bending needs to be taken
into account. This has a strong effect on the brightness of the
primary power-law source, making it appear faint to a distant
observer when it is close to the hole and bright when further
away. Some of the variability of the power-law continuum can thus
be due to the position of the source relative to the hole, rather than
any intrinsic effect. The strong light bending causes much of the
flux variability. 

\begin{figure}[h]
\centering
\includegraphics[width=0.65\linewidth,angle=270]{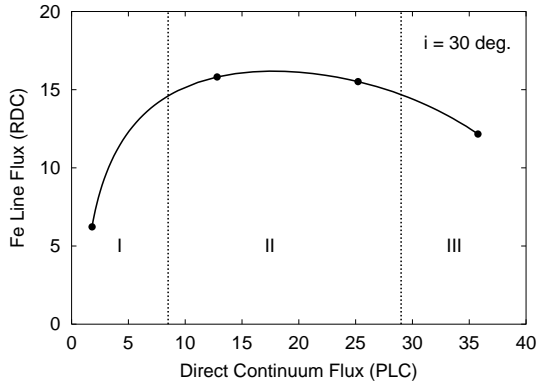}
\caption{The response of the reflection component (RDC) to the
power-law continuum (PLC) in the light bending model. The height of
the PLC above the hole is 20, 10, 5 and 1$r_{\rm g}$ at the 
4 marked points on the curve
going from right to left.}
\end{figure}

Consider a constant power-law source which is brought down the spin
axis from 20 to 1$r_{\rm g}$. It would appear to an observer seeing the
disc at an inclination of say 30 degrees to decrease dramatically in
flux. The reflection component would however change little until the
source is below about $4r_{\rm g}$. Although the reflection is
becoming more concentrated at the centre of the disc the increase in
power-law flux bent down onto the disc compensates for any loss of
flux. 

This 'light-bending' model (Miniutti et al 2003, 2004) is a simple
consequence of strong gravity close to the black hole and predicts
effects that have to be taken into account.

\begin{figure}[h]
\centering
\includegraphics[width=0.65\linewidth,angle=270]{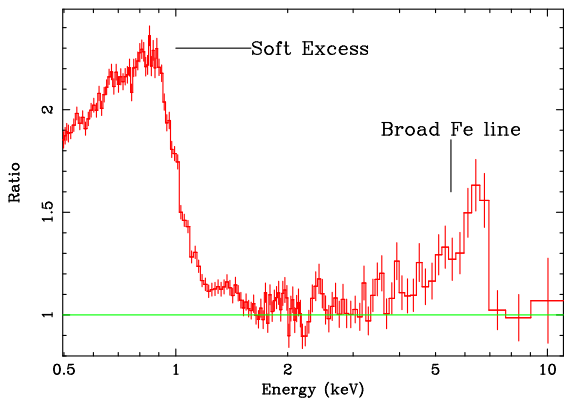}
\caption{The XMM-Newton spectrum of 1H\,0707 shown as the ratio to a
power-law fitted in the 2--3 and 8--10~keV energy bands. The spectral
features resemble blurred reflection, although absorption models can
be made to fit (Boller et al 2002). }
\end{figure}
\begin{figure}[h]
\centering
\includegraphics[width=0.65\linewidth,angle=270]{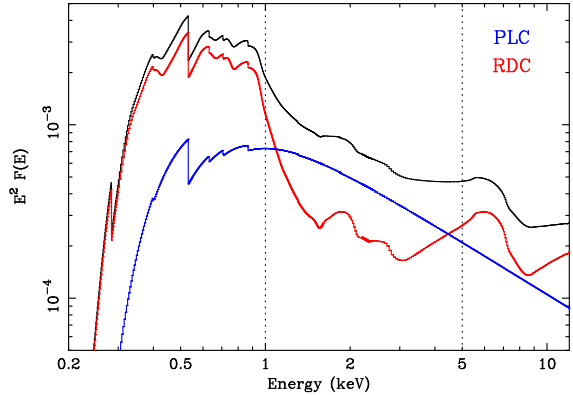}
\caption{The spectral components in the 2 component model for 1H\,0707
(Fabian et al 2002. In this state the power-law component dominates
the spectrum in the 1--5~keV band.
}
\end{figure}

\section{Applications to a range of sources}

The light-bending model has been applied to an increasing range of
 AGN, particularly NLS1 (MCG--6-30-15, Fabian \& Vaughan 2003;
 NGC\,4051, Ponti et al 2005; 1H0707, Fabian et al 2004; 1H0439,
 Fabian et al 2005) and at least one Galactic Black Hole ( GRO\,J1650,
 Rossi et al 2005). Future challenges are to see whether it fits just
 a class of AGN or its relevance is more widespread.

Is it consistent, for example, with the variable, red or blue-shifted,
emission features occasionally seen in some objects (e.g. NGC\,3516,
Iwasawa et al 2004)? A possibility in those cases is that indeed most
of the primary emission raises from close to the centre of the disc
but, due to the rapid rotation there, it is beamed along the disc and
illuminates transient `bumps' or waves on the surface of the disc,
causing transient reflection there.

What we need to do next is to see whether the reflection does follow
the variation expected to occur as the power-law continuum moves and
varies (Fig.~7). Subtle changes are expected in the degree of extreme
blurring which occurs when the power-law source is closest to the
black hole.  This requires more data from the parts of the lightcurve
when the flux is low. These occur infrequently but are accessible in
MCG--6-30-15 by long observations with XMM-Newton and Suzaku. Testable
variations of the reflection from the strong gravity regime around
black holes should be detectable now with long dedicated observations.

On the longer term we look to the next generation of detectors to
measure the reverberation of the reflection. This needs to be done in
a light crossing time and requires a large collecting area. AGN detect
100s of times more photons per light crossing time than Galactic black
holes so are the preferred targets. At a flux of about 2 photon per
square metre per second from the brightest iron lines, this requires a
collecting area of at least two square metres at 6~keV. We look forward
to such observations with Xeus and Con-X.

\section*{Acknowledgments}

I thank Giovanni Miniutti for help and discussion and the Royal
Society for support.

\end{document}